# P-CRE-DML: A Novel Approach for Causal Inference in Non-Linear Panel Data

Amarendra Sharma[1]

## Abstract

This paper introduces a novel Proxy-Enhanced Correlated Random Effects Double Machine Learning (P-CRE-DML) framework to estimate causal effects in panel data with non-linearities and unobserved heterogeneity. Combining Double Machine Learning (DML, Chernozhukov et al., 2018), Correlated Random Effects (CRE, Mundlak, 1978), and lagged variables (Arellano & Bond, 1991) and innovating within the CRE-DML framework (Chernozhukov et al., 2022; Clarke & Polselli, 2025; Fuhr and Papies, 2024), we apply P-CRE-DML to investigate the effect of social trust on GDP growth across 89 countries (2010–2020). We find positive and statistically significant relationship between social trust and economic growth. This aligns with prior findings on trust-growth relationship (e.g., Knack & Keefer, 1997). Furthermore, a Monte Carlo simulation demonstrates P-CRE-DML's advantage in terms of lower bias over CRE-DML and System GMM. P-CRE-DML offers a robust and flexible alternative for panel data causal inference, with applications beyond economic growth.

## 1. Introduction

This paper introduces Proxy-Enhanced Correlated Random Effects Double Machine Learning (LP-CRE-DML) framework, as a novel approach for causal inference in dynamic panel data. It integrates DML's causal inference (Chernozhukov et al., 2018),

---

[1] Department of Economics, Binghamton University, Binghamton, NY, USA. Email: aksharma@binghamton.edu

CRE's heterogeneity control (Mundlak, 1978), and proxied lagged variables (Arellano & Bond, 1991). We apply P-CRE-DML to estimate the causal effect of trust on GDP growth. Social trust has been postulated to promote economic growth by reducing transaction costs and boosting cooperation (Knack & Keefer, 1997; Zak & Knack, 2001). The Proxy-Enhanced Correlated Random Effects Double Machine Learning (P-CRE-DML) estimator is proposed as an innovation in panel data econometrics due to the inherent limitations of existing methods in the presence of dynamic confounding structures, including Ordinary Least Squares (OLS), Fixed Effects, Correlated Random Effects-Double Machine Learning (CRE-DML), and System Generalized Method of Moments (System GMM). OLS and Fixed Effects estimators often yield biased results in the presence of complex confounding, as they struggle to account for unobserved heterogeneity and endogeneity. CRE-DML, which builds on Mundlak (1978) and Chamberlain (1982), addresses time-invariant heterogeneity through time-averaged controls. However, it is susceptible to bias from time-varying unobserved confounders, such as policy shifts or cultural changes that simultaneously affect variables like trust and economic growth in our study (e.g., Chernozhukov et al., 2022; Clarke & Polselli, 2025; Fuhr and Papies, 2024). Analogously, System GMM (Blundell and Bond, 1998) uses lagged variables as instruments to address endogeneity and dynamic panel bias. But its reliance on the validity of these instruments weakens when time-varying confounders violate exclusion restrictions, which is further compounded by issues of multicollinearity. In contrast, P-CRE-DML integrates lagged proxies, such as prior-period government effectiveness, to capture dynamic confounders within a Double Machine Learning framework (Chernozhukov et al., 2018). It utilizes flexible machine learning models such as Random Forest to orthogonalize estimation and reduce multicollinearity. P-CRE-DML offers a robust and nuanced approach to causal inference by leveraging the temporal dynamics of panel data and relaxing stringent instrument assumptions. This feature makes it indispensable for economic analyses where unobserved heterogeneity evolves over time.

Using data from 89 countries from 2010-2020, we find statistically significant positive relationship between trust and growth at the conventional levels, confirming prior findings. We also perform a Monte Carlo simulation which confirms P-CRE-DML's lower bias and MSE compared to CRE-DML and System GMM. supporting its robustness in non-linear settings.

Section 2 details the methodology, Section 3 describes the data and presents empirical results, Section 4 presents simulation results, and Section 5 concludes.

## 2. Methodology

### 2.1 Double Machine Learning and Random Forest

We employ Double Machine Learning (DML) to estimate the causal effect of social trust on GDP growth. It is a framework that combines machine learning with econometric techniques to address high-dimensional confounding while maintaining valid inference. DML employs flexible machine learning models, such as Random Forest, to estimate nuisance parameters. This allows for robust estimation of treatment effects under partial linearity assumptions. This section outlines the DML framework, Random Forest model, CRE-DML, and P-CRE-DML.

### 2.2 Double Machine Learning Framework

DML, introduced by Chernozhukov et al. (2018), estimates the average treatment effect (ATE) of a treatment variable $D$ (e.g., social trust) on an outcome $Y$ (e.g., GDP growth) in the presence of high-dimensional confounders $X$ (e.g., capital, labor, technology, etc.). The model assumes a partially linear structure:

$$Y = \theta_0 D + g_0(X) + \epsilon, \quad E(\epsilon|D,X) = 0,$$

$$D = m_0(X) + v, \quad E(v|X) = 0,$$

where $\theta_0$ is the ATE of interest (causal effect of trust on GDP growth). $g_0(X)$ is a non-linear function capturing the effect of confounders on the outcome. $m_0(X)$ is a non-linear function modeling the treatment as a function of confounders. $\epsilon$ and $v$ are error terms with zero conditional mean. Direct estimation of $\theta_0$ using machine learning is biased due to overfitting of $g_0(X)$ and $m_0(X)$. DML mitigates this by using orthogonalized residuals and cross-fitting. The procedure is as follows:

Sample splitting involves partitioning the data into ($K$) folds (e.g., $K = 5$). For each fold ($K$), nuisance estimation is carried out by estimating $\hat{g}_0(X) = E(Y|X)$ and $\hat{m}_0(X) = E(D|X)$ on the out-of-fold data using a machine learning model (e.g., Random Forest). Next residuals are computed $\tilde{Y}_i = Y_i - \hat{g}_0(X_i)$, $\tilde{D}_i = D_i - \hat{m}_0(X_i)$. This is followed by Orthogonalized Regression, which involves estimating $\theta_0$ by regressing $\tilde{Y}_i$ on $\tilde{D}_I$ using the estimator $\hat{\theta}_0 = \left(\sum_{i=1}^{n} \tilde{D}_i^2\right)^{-1} \sum_{i=1}^{n} \tilde{D}_i \tilde{Y}_i$, The estimated $\hat{\theta}_0$ is cross fitted by averaging $\hat{\theta}_0$ across folds to reduce overfitting bias. Next, standard errors are computed for inference using the asymptotic variance:

$$Var(\theta_0) = \frac{1}{n} E\left[\frac{\tilde{\epsilon}_i^2 \tilde{D}_i^2}{\left(E[\tilde{D}_i^2]\right)^2}\right]$$

where $\tilde{\varepsilon}i = \tilde{Y}i - \hat{\theta}_0 \tilde{D}_i$. This ensures valid inference under weak assumptions on the machine learning estimators' convergence rates.

## 2.3 Random Forest as Nuisance Estimator

Random Forest (Breiman, 2001) is employed to estimate the nuisance functions $g_0(X)$ and $m_0(X)$ in the DML framework. Random Forest is an ensemble method that constructs multiple decision trees and aggregates their predictions. It offers robustness in the presence of non-linear relationships and high-dimensional data.

Formally, given a dataset $\{(Y_i, X_i)\}, \{i=1,...,n\}$, a Random Forest with $B$ trees predicts $\hat{f}(X)$ as: $\hat{f}(X) = \frac{1}{B}\sum_{b=1}^{B} T_b(X)$, where $T_b(X)$ is the prediction from the $b$-th tree. Each tree $T_b$ is constructed as follows:

Draw a random sample of size ($n$) with replacement from the data (bootstrap sample). At each node, select a random subset of $m$ features (typically $m = \sqrt{p}$ for $p$ features) to consider for splitting. This is known as feature subsampling. Next step involves Recursive Splitting. This implies splitting the node to minimize a loss function (e.g., mean squared error for regression):

$$MSE = \frac{1}{n_L}\sum_{i \in L}(Y_i - \bar{Y}_L)^2 + \frac{1}{n_R}\sum_{i \in R}(Y_i - \bar{Y}_R)^2,$$

where $L$ and $R$ are left and right child nodes, $n_L$ and $n_R$ are their sample sizes, and $\bar{Y}_L$, $\bar{Y}_R$ are the mean outcomes in each node. The stopping criteria requires continuing to split until a maximum depth or minimum node size is reached.

In our DML implementation, we use Random Forest Regressor from scikit-learn with $B = 50$ trees and *max depth* = *5*. Random Forest's ability to capture non-linear relationships makes it an effective choice for modeling $g_0(X)$ and $m_0(X)$, ensuring the residuals $\tilde{Y}_i$ and $\tilde{D}_I$ are orthogonalized for unbiased estimation of $\theta_0$.

## 2.4 Proxy-Enhanced Correlated Random Effects Double Machine Learning (P-CRE-DML)

The Proxy-Enhanced Correlated Random Effects Double Machine Learning (P-CRE-DML) estimator represents an evolution of econometric methodology. We develop it to overcome the limitations of standard Double Machine Learning (DML), which, despite its ability in accommodating high-dimensional covariates and non-linear relationships, treats data as cross-sectional and hence overlooks country-specific effects that may correlate with time-varying covariates, such as trust or capital in this study. Such correlations risk biasing causal estimates when unobserved factors—persistent institutional quality, cultural norms, or socio-political dynamics—influence both trust and GDP growth. P-CRE-DML extends the CRE-DML framework (Chernozhukov et al., 2022; Clarke & Polselli, 2025) by integrating lagged and mean proxy variables, specifically in the context of this paper, government effectiveness lag and government effectiveness mean, to address both time-varying and time-invariant unobserved confounders. This enhances robustness against dynamic endogeneity and multicollinearity.

In standard CRE-DML, time-invariant heterogeneity is mitigated through the inclusion of time-averaged treatment and outcome variables (e.g., trust mean and GDP growth mean in this study), as articulated by Fuhr and Papies (2024), within the model specifications

$$Y_{it} = \gamma g_0(X_{it}, \bar{X}, \bar{D}_i) + \mu_{it}$$

and

$$D_{it} = \partial m_0(X_{it}, \bar{X}_i, \bar{D}_i) + \eta_{it}.$$

This approach effectively controls for stable, unit-specific unobserved effects without inflating parameter counts. However, panel data frequently exhibit dynamic endogeneity, where lagged outcomes $Y_{i,t-1}$ or treatments $D_{i,t-1}$ influence current values and potentially correlate with the contemporaneous error term $\mu_{it}$. P-CRE-DML innovatively incorporates lagged proxy variables, such as government effectiveness lag, which serve as instrumental variables satisfying relevance (correlated with outcome $Y_{it}$ and treatment $D_{it}$ and exogeneity (uncorrelated with contemporaneous shocks $\epsilon_{it}$, conditional on controls). This lagged proxy captures time-varying confounders—policy shocks, institutional transitions, or socio-political shifts—that dynamically affect trust and economic growth outcomes, leveraging the temporal structure of panel data to satisfy the proxy variable assumption, thus mitigating biases unaddressed by standard CRE-DML.

Complementing this, the inclusion of government effectiveness mean aligns with the CRE framework's logic, capturing time-invariant unobserved heterogeneity, such as stable governance structures or institutional stability, which may correlate with government effectiveness and confound the trust-growth relationship. By incorporating this time-averaged proxy, P-CRE-DML isolates the causal effect of the endogenous variable, neutralizing biases from persistent, unobserved unit-specific factors. The inclusion of government effectiveness lag and government effectiveness mean within the Double Machine Learning framework reinforces P-CRE-DML's resilience against multicollinearity and misspecification, offering a robust tool for causal inference in economic panel data marked by complex, evolving confounding dynamics.

The general form of the partially linear regression (PLR) model in P-CRE-DML can be written as:

$$Y_{it} = \theta_0 D_{it} + g_0(X_{it}, \bar{X}_\iota, Z_{i,t-1}, \alpha_i) + \epsilon_{it}$$

where:

$Y_{it}$: Outcome variable for unit $i$ at time $t$, here GDP growth.

$D_{it}$: Treatment variable, here trust.

$\theta_0$: Causal parameter of interest, representing the effect of trust on GDP growth.

$X_{it}$: Vector of observed time-varying control variables (e.g., capital, labor, technology).

$\bar{X}_\iota$: Vector of unit-specific means of time-varying variables (CRE terms, e.g., trust mean, GDP growth mean, government effectiveness mean), capturing unobserved heterogeneity correlated with regressors.

$Z_{i,t-1}$: Vector of lagged proxy variables (e.g., GDP growth lag, government effectiveness lag), addressing time-varying confounders.

$\alpha_i$: Unobserved unit-specific heterogeneity (partially absorbed by CRE terms).

$g_0$: Unknown non-linear function capturing the effects of controls, CRE terms, and lagged proxies.

$\epsilon_{it}$: Idiosyncratic error term, assumed to satisfy $E(\epsilon_{it}|D_{it}, X_{it}, \bar{X}_\iota, Z_{i,t-1}, \alpha_i) = 0$.

Additionally, the treatment variable is modeled as:

$$D_{it} + m_0\left(X_{it}, \bar{X}_i, Z_{i,t-1}, \alpha_i\right) + v_{it}$$

where $m_0$ is an unknown function, and $v_{it}$ is an error term with $E(v_{it}| X_{it}, \bar{X}_i, Z_{i,t-1}, \alpha_i) = 0$.

The P-CRE-DML framework uses DML to estimate $\theta_0$ by partialling out the effects of $g_0$ and $m_0$, leveraging machine learning (here, Random Forest Regressors) to flexibly model these functions while incorporating CRE terms to control for unobserved heterogeneity and lagged proxies to address dynamic confounding.

**2.4 Implementation**

In our empirical analysis, we use DML to estimate the effect of social trust $D$ on GDP growth $Y$, controlling for confounders $X$ = {capital, labor, technology}. The DML estimator is implemented via the DoubleMLPLR class in the DoubleML Python package, with $K$ = 5 folds and one repetition. System GMM is implemented via moment conditions minimized using scipy.optimize, with bootstrap standard errors. The python codes used in this paper are provided in the appendix.

**3. Data and Empirical Results**

**3.1 Data Sources and Description**

Our analysis draws on a panel dataset of 89 countries over the period 2010–2020 comprising 944 observations for the full sample. It includes countries with at least 8 years of technology data over 2010–2020, ensuring sufficient temporal coverage for panel analysis. Missing values were imputed via linear interpolation within country panels, supplemented by mean imputation for residual gaps.

Data are sourced from multiple repositories:

World Bank World Development Indicators (WDI): Provides economic variables such as GDP per capita growth, capital formation, labor participation, and Research & Development spending.

Worldwide Governance Indicators (WGI): Supplies the government effectiveness index.

World Values Survey (WVS): Offers the trust measure based on responses to the question, "Can most people be trusted?"

Table 1 below presents the estimation results of all estimators.

**Table 1**

| Estimator | Coefficient | Standard Error | p-value |
|---|---|---|---|
| OLS | 0.0253 | 0.0074 | 0.0007 |
| Fixed Effects | 0.0338 | 0.0175 | 0.0544 |
| System GMM | 0.1642 | 0.0616 | 0.0077 |
| CRE-DML | 0.0573 | 0.0256 | 0.0255 |
| P-CRE-DML | 0.0592 | 0.0241 | 0.0139 |

The trust coefficient ranges from 0.0253 (OLS) to 0.1642 (System GMM). The trust coefficients for CRE-DML and P-CRE-DML are 0.0573 and 0.0592, respectively. The corresponding *p*-values are 0.0255 and 0.0139. This clearly shows that P-CRE-DML estimate has higher statistical significance than the CRE-DML estimator.

## 4. Simulation Study

We conducted a Monte Carlo simulation (1000 iterations, N=89, T=11) with the DGP:

$$y_{it} = 0.3 y_{i,t-1} + 0.1 d_{it} + \sum_{j=1}^{k} x_{j,it} + 0.05 d_{it}^2 + \alpha_i + \varepsilon_{it}$$

where $d_{it} = 0.5v_{it} + 0.3\alpha_i$ includes a non-linear trust effect ($\theta$=0.1). Results (Table 2) show P-CRE-DML's bias (0.1682) and MSE (0.0461) are significantly lower than CRE-DML's (0.3279, 0.1253), but the variances are same (0.0178). P-CRE-DML outperforms System GMM. The bias under System GMM is 5.4572 and MSE is 627.7925, which are significantly higher than the P-CRE-DML's bias and MSE. Figure 1 displays bias estimate distributions, with P-CRE-DML clustering around $\theta$=0.16.

**Table 2: Simulation Results**

| Estimator | Bias | Variance | MSE |
|---|---|---|---|
| OLS | 3.5723 | 0.1424 | 12.9037 |
| Fixed Effects | 0.0981 | 0.0089 | 0.0185 |
| System GMM | 5.4572 | 598.0111 | 627.7925 |
| CRE-DML | 0.3279 | 0.0178 | 0.1253 |
| P-CRE-DML | 0.1682 | 0.0178 | 0.0461 |

**Figure 1: Histogram of Simulation Bias**

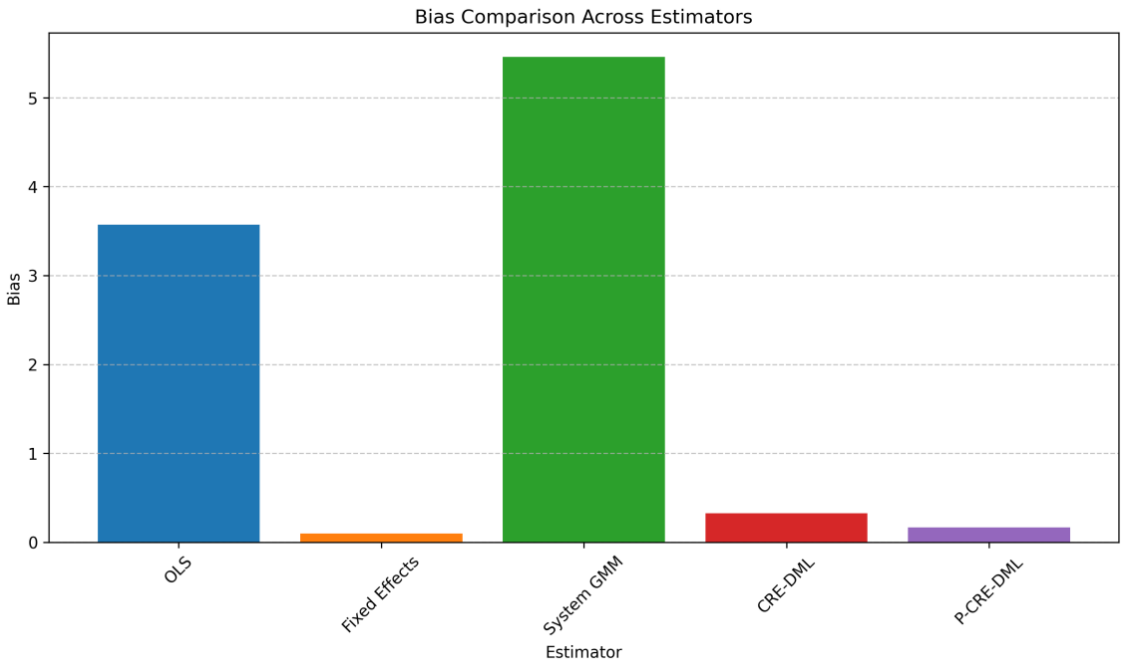

P-CRE-DML's flexibility in modeling non-linearities and controlling heterogeneity via CRE and lagged instrument terms makes it a robust alternative to System GMM and CRE-DML, especially in short panels with potential instrument weaknesses.

## 5. Conclusion

This study introduces P-CRE-DML estimator by incorporating lagged and mean values of proxy variables that are assumed to be correlated with the outcome and treatment variables in a non-linear panel data setting within the CRE-DML framework. We demonstrate its advantages over CRE-DML and System GMM in non-linear, short-panel settings via Monte Carlo simulation. Additionally, using this estimator, we find positive and statistically significant trust-growth effect. The statistical significance of estimated trust coefficient is greater in the case of P-CRE-DML relative to CRE-DML. P-CRE-DML's flexibility and robustness makes it a valuable tool for panel data causal inference, with applications in economics and beyond.

# Appendix

**Python Code to implement P-CRE-DML and other estimators used in this paper**

```
import pandas as pd
import numpy as np
from doubleml import DoubleMLData, DoubleMLPLR
from sklearn.ensemble import RandomForestRegressor
from scipy.optimize import minimize
from statsmodels.regression.linear_model import OLS
from statsmodels.tools.tools import add_constant
from statsmodels.stats.outliers_influence import variance_inflation_factor
import matplotlib.pyplot as plt
from sklearn.inspection import partial_dependence
import seaborn as sns
```

```python
from scipy.stats import norm
from tqdm import tqdm
from linearmodels.panel import PanelOLS
import os

# Set random seed for reproducibility
np.random.seed(42)

# --- Real Data Analysis ---

# OLS
def run_ols(df):
    X = df[['trust', 'capital', 'labor', 'technology', 'gov_effectiveness_lag']]
    X = add_constant(X)
    model = OLS(df['gdp_growth'], X).fit()
    coef = model.params['trust']
    se = model.bse['trust']
    p_value = model.pvalues['trust']
    return {'coef': coef, 'se': se, 'p_value': p_value}
# Fixed Effects
def run_fixed_effects(df):
    df_fe = df.set_index(['country', 'year'])
    exog = df_fe[['trust', 'capital', 'labor', 'technology', 'gov_effectiveness_lag']]
    model = PanelOLS(df_fe['gdp_growth'], exog, entity_effects=True).fit()
    coef = model.params['trust']
    se = model.std_errors['trust']
    p_value = model.pvalues['trust']
    return {'coef': coef, 'se': se, 'p_value': p_value}

# CRE-DML (original, with trust_mean and gdp_growth_mean)
def run_cre_dml(df, subsample_name='full'):
    control_vars = ['capital', 'labor', 'technology', 'gdp_growth_lag', 'trust_mean', 'gdp_growth_mean']
    dml_data = DoubleMLData(df, y_col='gdp_growth', d_cols=['trust'], x_cols=control_vars)
    dml_rf = DoubleMLPLR(dml_data,
                ml_l=RandomForestRegressor(n_estimators=200, max_depth=15, min_samples_split=5, random_state=42),
                ml_m=RandomForestRegressor(n_estimators=200, max_depth=15, min_samples_split=5, random_state=42),
                n_folds=5, n_rep=1, score='partialling out')
    dml_rf.fit()
    theta = dml_rf.coef[0]
    se = dml_rf.se[0]
    p_value = 2 * (1 - norm.cdf(abs(theta / se)))
```

```python
# P-CRE-DML (with lagged proxy for time-varying confounders)
def run_p_cre_dml(df, subsample_name='full'):
    control_vars = ['capital', 'labor', 'technology', 'gdp_growth_lag',
'gov_effectiveness_lag', 'trust_mean', 'gdp_growth_mean', 'gov_effectiveness_mean']
    dml_data = DoubleMLData(df, y_col='gdp_growth', d_cols=['trust'], x_cols=control_vars)
    dml_rf = DoubleMLPLR(dml_data,
                ml_l=RandomForestRegressor(n_estimators=200, max_depth=15, min_samples_split=5, random_state=42),
                ml_m=RandomForestRegressor(n_estimators=200, max_depth=15, min_samples_split=5, random_state=42),
                n_folds=5, n_rep=1, score='partialling out')
    dml_rf.fit()
    theta = dml_rf.coef[0]
    se = dml_rf.se[0]
    p_value = 2 * (1 - norm.cdf(abs(theta / se)))

# System GMM
def run_system_gmm(df, subsample_name='full'):
    df = df.copy()
    df['trust_lag2'] = df.groupby('country')['trust'].shift(2)
    df['trust_lag3'] = df.groupby('country')['trust'].shift(3)
    df['gdp_growth_lag2'] = df.groupby('country')['gdp_growth'].shift(2)
    df['gdp_growth_lag3'] = df.groupby('country')['gdp_growth'].shift(3)
    df = df.dropna()

    def moment_conditions(params, df):
        rho, beta, gamma1, gamma2, gamma3, gamma4 = params
        y = df['gdp_growth'].values
        X = df[['gdp_growth_lag', 'trust', 'capital', 'labor', 'technology', 'gov_effectiveness_lag']].values
        Z = df[['gdp_growth_lag2', 'trust_lag2', 'trust_lag3']].values
        errors = y - X @ params
        moments = Z.T @ errors
        return np.mean(moments**2)

    initial_params = np.zeros(6)
    result = minimize(moment_conditions, initial_params, args=(df,), method='Nelder-Mead')
    params = result.x
    beta = params[1]

    n_boot = 10
    boot_betas = []
    for _ in range(n_boot):
```

```python
        boot_df = df.sample(frac=1, replace=True)
        try:
            boot_result = minimize(moment_conditions, initial_params, args=(boot_df,), method='Nelder-Mead')
            boot_betas.append(boot_result.x[1])
        except:
            continue
    se = np.std(boot_betas) if boot_betas else np.nan
    p_value = 2 * (1 - norm.cdf(abs(beta / se))) if not np.isnan(se) else np.nan

    return {'coef': beta, 'se': se, 'p_value': p_value}

# Run estimators
try:
    ols_results = run_ols(df)
except KeyError as e:
    print(f"OLS failed: {e}")
    ols_results = {'coef': np.nan, 'se': np.nan, 'p_value': np.nan}

try:
    fe_results = run_fixed_effects(df)
except KeyError as e:
    print(f"Fixed Effects failed: {e}")
    fe_results = {'coef': np.nan, 'se': np.nan, 'p_value': np.nan}

try:
    cre_dml_results = run_cre_dml(df)
except KeyError as e:
    print(f"CRE-DML failed: {e}")
    cre_dml_results = {'coef': np.nan, 'se': np.nan, 'p_value': np.nan}

try:
    p_cre_dml_results = run_p_cre_dml(df)
except KeyError as e:
    print(f"P-CRE-DML failed: {e}")
    p_cre_dml_results = {'coef': np.nan, 'se': np.nan, 'p_value': np.nan}

try:
    gmm_results = run_system_gmm(df)
except KeyError as e:
    print(f"System GMM failed: {e}")
    gmm_results = {'coef': np.nan, 'se': np.nan, 'p_value': np.nan}

# Real data results table
real_results_df = pd.DataFrame({
    'Estimator': ['OLS', 'Fixed Effects', 'System GMM', 'CRE-DML', 'P-CRE-DML'],
```

```python
        'Coefficient': [ols_results['coef'], fe_results['coef'], gmm_results['coef'],
cre_dml_results['coef'], p_cre_dml_results['coef']],
        'Standard Error': [ols_results['se'], fe_results['se'], gmm_results['se'],
cre_dml_results['se'], p_cre_dml_results['se']],
        'p-value': [ols_results['p_value'], fe_results['p_value'], gmm_results['p_value'],
cre_dml_results['p_value'], p_cre_dml_results['p_value']]
})
real_results_df = real_results_df.round(4)
real_results_df.to_csv('/Users/your ID/Desktop/real_data_results.csv', index=False)
print("\nReal Data Results Table:\n", real_results_df)
print("Saved to: /Users/your ID/Desktop/real_data_results.csv")

# --- Simulation Analysis ---

# Simulate panel data with time-varying confounder
N, T = 89, 11
n_obs = N * T
countries = np.repeat(np.arange(N), T)
years = np.tile(np.arange(T), N)
true_theta = 0.1
n_simulations = 1000
ols_results_sim = []
fe_results_sim = []
cre_dml_results_sim = []
p_cre_dml_results_sim = []
gmm_results_sim = []

for sim in tqdm(range(n_simulations), desc="Simulations"):
    alpha_i = np.repeat(np.random.normal(0, 1, N), T)
    x_it = np.random.normal(0, 1, (n_obs, 3))
    proxy_it = np.random.normal(0, 1, n_obs)  # Proxy for time-varying confounder
    proxy_lag = np.roll(proxy_it, T)
    proxy_lag[:T] = 0
    alpha_it = 0.5 * np.roll(alpha_i, 1) + 0.3 * proxy_lag + np.random.normal(0, 0.2, n_obs)  # Time-varying confounder
    alpha_it[:T] = 0
    d_it = 0.2 * np.random.normal(0, 1, n_obs) + 0.1 * alpha_i + 0.1 * alpha_it
    d_lag = np.roll(d_it, T)
    d_lag[:T] = 0
    y_it = np.zeros(n_obs)
    for t in range(1, T):
        idx = np.where(years == t)[0]
        y_it[idx] = 0.3 * y_it[idx - T] + true_theta * d_it[idx] + x_it[idx].sum(axis=1) + \
            0.05 * d_it[idx]**2 + alpha_i[idx] + alpha_it[idx] + np.random.normal(0, 0.5, len(idx))
    df_sim = pd.DataFrame({
```

```python
        'country': countries, 'year': years, 'gdp_growth': y_it, 'trust': d_it,
        'trust_lag': d_lag, 'gdp_growth_lag': np.roll(y_it, T),
        'capital': x_it[:, 0], 'labor': x_it[:, 1], 'technology': x_it[:, 2],
        'gov_effectiveness_lag': proxy_lag
    })

    # Create mean variables
    mean_vars_sim = df_sim.groupby('country')[['trust', 'gdp_growth',
'gov_effectiveness_lag']].mean().reset_index()
    mean_vars_sim = mean_vars_sim.rename(columns={'trust': 'trust_mean',
'gdp_growth': 'gdp_growth_mean', 'gov_effectiveness_lag': 'gov_effectiveness_mean'})
    df_sim = df_sim.merge(mean_vars_sim, on='country')

    df_sim.loc[:T-1, 'gdp_growth_lag'] = 0
    df_sim = df_sim[df_sim['year'] >= 1]

    # Run estimators
    try:
        ols_result = run_ols(df_sim)
        ols_results_sim.append(ols_result['coef'])
    except Exception as e:
        print(f"Simulation {sim} OLS failed: {e}")
        ols_results_sim.append(np.nan)

    try:
        fe_result = run_fixed_effects(df_sim)
        fe_results_sim.append(fe_result['coef'])
    except Exception as e:
        print(f"Simulation {sim} Fixed Effects failed: {e}")
        fe_results_sim.append(np.nan)

    try:
        cre_dml_result = run_cre_dml(df_sim)
        cre_dml_results_sim.append(cre_dml_result['coef'])
    except Exception as e:
        print(f"Simulation {sim} CRE-DML failed: {e}")
        cre_dml_results_sim.append(np.nan)

    try:
        p_cre_dml_result = run_p_cre_dml(df_sim)
        p_cre_dml_results_sim.append(p_cre_dml_result['coef'])
    except Exception as e:
        print(f"Simulation {sim} P-CRE-DML failed: {e}")
        p_cre_dml_results_sim.append(np.nan)

    try:
```

```python
        gmm_result = run_system_gmm(df_sim)
        gmm_results_sim.append(gmm_result['coef'])
    except Exception as e:
        print(f"Simulation {sim} System GMM failed: {e}")
        gmm_results_sim.append(np.nan)

# Simulation metrics
ols_results_sim = [x for x in ols_results_sim if not np.isnan(x)]
fe_results_sim = [x for x in fe_results_sim if not np.isnan(x)]
cre_dml_results_sim = [x for x in cre_dml_results_sim if not np.isnan(x)]
p_cre_dml_results_sim = [x for x in lp_cre_dml_results_sim if not np.isnan(x)]
gmm_results_sim = [x for x in gmm_results_sim if not np.isnan(x)]

ols_bias = np.mean(ols_results_sim) - true_theta if ols_results_sim else np.nan
fe_bias = np.mean(fe_results_sim) - true_theta if fe_results_sim else np.nan
cre_dml_bias = np.mean(cre_dml_results_sim) - true_theta if cre_dml_results_sim else np.nan
p_cre_dml_bias = np.mean(lp_cre_dml_results_sim) - true_theta if lp_cre_dml_results_sim else np.nan
gmm_bias = np.mean(gmm_results_sim) - true_theta if gmm_results_sim else np.nan

ols_var = np.var(ols_results_sim) if ols_results_sim else np.nan
fe_var = np.var(fe_results_sim) if fe_results_sim else np.nan
cre_dml_var = np.var(cre_dml_results_sim) if cre_dml_results_sim else np.nan
p_cre_dml_var = np.var(lp_cre_dml_results_sim) if lp_cre_dml_results_sim else np.nan
gmm_var = np.var(gmm_results_sim) if gmm_results_sim else np.nan

ols_mse = np.mean((np.array(ols_results_sim) - true_theta)**2) if ols_results_sim else np.nan
fe_mse = np.mean((np.array(fe_results_sim) - true_theta)**2) if fe_results_sim else np.nan
cre_dml_mse = np.mean((np.array(cre_dml_results_sim) - true_theta)**2) if cre_dml_results_sim else np.nan
p_cre_dml_mse = np.mean((np.array(lp_cre_dml_results_sim) - true_theta)**2) if lp_cre_dml_results_sim else np.nan
gmm_mse = np.mean((np.array(gmm_results_sim) - true_theta)**2) if gmm_results_sim else np.nan

# Simulation results table
sim_results_df = pd.DataFrame({
    'Estimator': ['OLS', 'Fixed Effects', 'System GMM', 'CRE-DML', 'P-CRE-DML'],
    'Bias': [ols_bias, fe_bias, gmm_bias, cre_dml_bias, p_cre_dml_bias],
    'Variance': [ols_var, fe_var, gmm_var, cre_dml_var, p_cre_dml_var],
    'MSE': [ols_mse, fe_mse, gmm_mse, cre_dml_mse, p_cre_dml_mse]
})
sim_results_df = sim_results_df.round(4)
```

```python
sim_results_df.to_csv('/Users/ your ID /Desktop/simulations_results_table.csv',
index=False)
print("\nSimulation Results Table:\n", sim_results_df)
print("Saved to: /Users/your ID/Desktop/simulations_results_table.csv")

# Simulation bias graph
plt.figure(figsize=(12, 8))
if ols_results_sim:
    sns.histplot(ols_results_sim, label='OLS', color='red', alpha=0.4, bins=20)
if fe_results_sim:
    sns.histplot(fe_results_sim, label='Fixed Effects', color='green', alpha=0.4, bins=20)
if gmm_results_sim:
    sns.histplot(gmm_results_sim, label='System GMM', color='orange', alpha=0.4,
bins=20)
if cre_dml_results_sim:
    sns.histplot(cre_dml_results_sim, label='CRE-DML', color='blue', alpha=0.4, bins=20)
if p_cre_dml_results_sim:
    sns.histplot(p_cre_dml_results_sim, label='P-CRE-DML', color='purple', alpha=0.4,
bins=20)
plt.axvline(true_theta, color='black', linestyle='--', label='True $\\theta = 0.1$')
plt.xlabel('Estimated Trust Coefficient')
plt.ylabel('Frequency')
plt.title('Distribution of Estimated Trust Coefficients Across Models')
plt.legend()
plt.savefig('/Users/your ID/Desktop/simulation_bias_graph.png', dpi=300)
plt.close()
print("Simulation bias graph saved to: /Users/your
ID/Desktop/simulation_bias_graph.png")
```